\documentclass[manuscript]{aastex}

\newcommand{\HST}{{\it HST}}
\newcommand{\IUE}{{\it IUE}}

\begin{document}                                                                     

\title{Binary Cepheids: Separations and Mass Ratios in $5\,M_\odot$
Binaries\altaffilmark{1}
}


\author{Nancy Remage Evans\altaffilmark{2,3},
Howard E. Bond\altaffilmark{4,5},
Gail H. Schaefer\altaffilmark{6},
Brian D. Mason\altaffilmark{7}, 
Margarita Karovska\altaffilmark{2},  
and   
Evan Tingle\altaffilmark{2}  
}  

\altaffiltext{1}
{Based in part on observations made with the NASA/ESA {\it Hubble Space
Telescope}, obtained by the Space Telescope Science Institute. STScI is operated
by the Association of Universities for Research in Astronomy, Inc., under NASA
contract NAS5-26555.}

\altaffiltext{2}
{Smithsonian Astrophysical Observatory,    
MS 4, 60 Garden St., Cambridge, MA 02138; nevans@cfa.harvard.edu}

\altaffiltext{3}
{Guest Observer with the {\it International Ultraviolet Explorer}, operated by
the Goddard Space Flight Center, National Aeronautics and Space Administration.}

\altaffiltext{4}
{Dept.\ of Astronomy \& Astrophysics, Pennsylvania State University,
University Park, PA 16802; heb11@psu.edu}

\altaffiltext{5}
{Space Telescope Science Institute, 3700 San Martin Drive, Baltimore, MD
21218}

\altaffiltext{6} {The CHARA Array, Georgia State University, P.O. Box 3965,
Atlanta GA 30302-3965; schaefer@chara-array.org}

\altaffiltext{7}{US Naval Observatory, 3450 Massachusetts Ave., NW, 
Washington, DC 20392-5420 }
 view

\begin{abstract}

Deriving the distribution of binary parameters for a particular class of stars
over the full range of orbital separations usually requires the combination of
results from many different observing techniques (radial velocities,
interferometry, astrometry, photometry, direct imaging), each  with selection
biases. However, Cepheids---cool, evolved stars of $\sim$$5\, M_\odot$---are a
special case because ultraviolet spectra will immediately reveal any companion
star hotter than early type~A, {\it regardless of the orbital separation}. We
have used {\it  International Ultraviolet Explorer\/} (\IUE\/) UV spectra of a
complete sample of all 76 Cepheids brighter than $V=8$ to create a list of {\it
all 18\/} Cepheids with companions more massive than $2.0\, M_\odot$.   Orbital
periods of many of these binaries are available from radial-velocity studies, or
can be estimated for longer-period systems from detected velocity variability.
In an imaging survey with the {\it Hubble Space Telescope\/} Wide Field
Camera~3, we resolved three of the companions (those of $\eta$ Aql, S Nor, and
V659 Cen), allowing us to make estimates of the periods out to the long-period
end of the distribution. Combining these  separations with orbital data in the
literature, we derive an unbiased  distribution of binary separations, orbital
periods, and mass ratios.   The distribution of orbital periods shows that the
$5\, M_\odot$ binaries have systematically shorter  periods than do $1\,
M_\odot$ stars. Our data also suggest  that  the distribution of mass ratios 
depends both on binary separation and system multiplicity.  The distribution of
mass ratios as a function of orbital separation, however, does not depend on
whether a system is a binary or a triple.

\end{abstract}


\keywords{binaries: general, stars: massive, stars: variables: Cepheids}


\section{Introduction}

Binary-star studies are valuable for what they provide directly (e.g., stellar
masses), as well as for the information they provide about the configurations 
resulting from star-formation processes.  This topic was particularly well
developed in a classical series of studies by Abt and collaborators.  For
instance, Abt, Gomez, \& Levy (1990) discussed this question for late B stars.

For several decades, binary-star studies  have been the beneficiary of
developments in  observational techniques, particularly those providing high
spatial resolution and  access to new wavelength regions.  A clear demonstration
of  progress  in this area was the discussion by Duquennoy \& Mayor (1991) of
the binary  properties of solar-mass stars.  They combined extensive CORAVEL
radial-velocity (RV) observations with results from visual binaries and
common-proper-motion stars  to explore distributions of mass ratios and
eccentricities at all separations.   Recently this work has been updated by
Raghavan et al.\ (2010) to  include new advances in high-resolution techniques
(long-baseline interferometry and speckle interferometry).  Stars more massive
than solar type are more difficult to study, because  they are rarer, and hence
more distant, and also because they have broad spectral  lines, which limit the
accuracy of RVs.  However, new observational techniques have likewise greatly
enhanced the knowledge of their properties (e.g., Kobulnicky \& Fryer 2007;
Kouwenhoven et al.\ 2007; Mason et al.\ 2009; Sana \& Evans 2011).  Sana \&
Evans, for instance, find a fairly constant fraction (44\%) of spectroscopic
binaries among OB stars in several nearby open clusters.   Systems with small
mass ratios (i.e., low secondary masses) are the most difficult to identify. 
Evans et al.\  (2011a) have used a  different approach to determine the fraction
of B stars with low-mass companions.  Since late B stars  produce X-rays
very rarely, the fraction of late B stars in the young cluster Tr~16 (associated
with $\eta$~Car) that were detected in X-rays provides the fraction (32--39\%)
which have young low-mass companions.  

Comparing the observed properties of binary and multiple systems with
star-formation  calculations is a test of the model predictions.  An obvious
first step of this approach  is a comparison of the properties of binary systems
containing high- and low-mass primary stars, but our knowledge of binaries among
intermediate- and high-mass stars is still not as extensive as it is for
solar-mass stars. 

This paper is the first in a series aimed at determining the properties of
binary systems containing Cepheid variables.  Cepheids are stars of intermediate
masses, ranging from about 4 to $9\,M_\odot$; in this paper we will use
$5\,M_\odot$ as the typical Cepheid mass. Cepheids are particularly useful for
determining binary properties for several reasons.  They have narrow spectral
lines, providing accurate RVs from optical spectroscopy.  If a Cepheid has a
fairly high-mass companion, the companion will dominate  the light of the system
in the ultraviolet (UV), thus immediately demonstrating that the system is a
binary.  This further makes it possible to determine masses by measuring the
orbital velocity amplitude of the companion in the  UV, for example by using
{\it Hubble Space Telescope\/} (\HST) spectra (e.g., Evans et al.\ 2011b). The
combination of the optical and UV RV curves provides the mass ratio, and if the
mass of the hot companion is inferred from its spectral type, the actual mass of
the Cepheid.  Such studies provide direct evidence on the distribution of  mass
ratios in binaries containing Cepheids (e.g., Evans 1995). They also provide
information about the fraction of triple systems (Evans et al.\ 2005), because
the companions can be directly studied in the UV\null.  A number of
Cepheid-containing triple systems have been identified through RV variability of
the companions (or inferred from the orbital mass functions). As compared with a
sample of single-lined spectroscopic binaries, the ability to directly observe
the companions provides a much higher detection rate of triples. 

This paper focuses on a complete sample of B- and early A-type companions of Cepheid
variables, which was obtained through a survey with the {\it International
Ultraviolet Explorer\/} (\IUE) satellite, as described in \S2. This approach has
the strength that the survey is sensitive to binary companions at all possible
separations. By contrast, RV studies only find the close systems. Conversely,
the limitation of this approach is that it does not detect low-mass companions.

The properties of the massive companion set include a few results from
our  recent  \HST\/ snapshot imaging survey of
Cepheids with the Wide Field Camera 3 (WFC3)---to be described in more detail in
a subsequent paper---as well as orbital information on our sample from  the
literature.  In the following sections we discuss the construction of the 
sample, the derivation of the orbital separations and mass ratios,  their
distribution functions, and some implications of our results.

\section{The Sample}

In order to have  a well-defined sample, we start with a spectroscopic survey of
all 76  Galactic Cepheids brighter than visual magnitude 8, which was
carried out with the \IUE\/ satellite by one of us (Evans 1992a). These spectra,
obtained with \IUE's LWP and LWR cameras, covered the near-ultraviolet (NUV)
wavelength range 2000 to 3200~\AA\null. From this study we selected the Cepheids
for which \IUE\/ revealed a companion of spectral type A2~V or earlier,
corresponding to companion masses greater than about $2\,M_\odot$ (e.g.,
Harmanec 1988). These are highly probable physical companions, because the
rarity of A- and B-type stars in the field makes it very unlikely that such a
star would be within the  \IUE\/ aperture by chance.

The sensitivity of this \IUE\/ survey  to hot companions varies somewhat from
star to star,  because of differences in the intrinsic luminosity of the
Cepheids, differences in the pulsation phases (hence magnitude and color) at the
time of the observation, and different exposure times.  One product of the
survey was a list of the spectral types of the brightest companions of each
Cepheid  that would {\it not\/} have been detected (Evans 1992a,
Table~1C)\null.  These limits were generally mid-A spectral types, but they were
early A for some, and late~B for four stars.  

Of the complete sample of 76 Cepheids observed in the NUV with \IUE, 15 of them
had detected A- or B-type companions (Evans 1992a, Table~3A)\null.  To this list
we have added three more Cepheids: (1)~V636~Sco and T~Vul, because hot
companions of both stars were detected with the \IUE\/ far-ultraviolet (FUV)
spectrograph and SWP camera (Evans 1992b, Table~3B).  They
were not evident on the 2000 to 3200~\AA\null\/ exposures presumably
because  of the phase of the Cepheid. (2)~$\delta$ Cep
itself, because trigonometric parallaxes obtained with the Fine Guidance System
(FGS) on \HST\/ showed that the Cepheid and its $40''$ B-type companion HD
213307 are at the same distance (Benedict et al.\ 2002).  Our list of the 18
Cepheids with companions of $2\, M_\odot$ or more is given in Table~1.

For each Cepheid that had a detected hot companion, spectral types are available
from \IUE\/ observations in the FUV spectral range (1150--1950~\AA)\null. FUV
spectra of late B and early A stars are particularly sensitive to temperature
changes, and such companions completely dominate the spectra, so the spectral
types are tightly constrained. The spectral classifications were derived by
comparison with \IUE\/ SWP spectra of MK standard stars.  Because the spectral
energy distribution is so temperature-sensitive, many of the companions were
found to have spectral types lying between those of the MK standards, resulting
in fractional spectral types such as B9.8~V\null.   Table~1 lists the  spectral
types for the  companions and the references  from which they were taken (cols.\
2 and 3). Many of the cited sources provide plots directly comparing the 
companion spectra with those of MK standards.  

Masses of the companions were derived from their spectral types, and are given
in col.~4 of Table~1.  For the late B and early A companions, the large
luminosity difference between the Cepheid and the companion means that the
companion can be assumed to lie on the zero-age main sequence (ZAMS)\null. For
the ZAMS, we use the Harmanec (1988) calibration of masses vs.\ spectral types.
The masses for these stars in Table~1 are mostly taken from Evans (1995), but
are new values for $\eta$~Aql, SU~Cas, S~Nor, T~Vul, and the systems discussed
below.

Table~1 also contains three hotter  companions, some of which
may  have evolved
beyond the ZAMS; these objects were discussed by Evans (1994).  Masses for AX
Cir, BP Cir (overtone mode), and V659 Cen are from Fig.~7 in that paper, and are
based on the Geneva evolutionary tracks, which include mild core convective
overshoot.  

A few companion masses in Table 1 require further discussion.  

(1)~{\it
$\delta$~Cep}: The spectral type and mass for the  companion, HD~213307, are
taken from Benedict et al.\ (2002). 

(2)~{\it S~Mus}: The companion spectral type
was derived  from {\it Far Ultraviolet Spectroscopic Explorer\/} ({\it FUSE})
spectra (Evans et al.\ 2006), from which the mass was derived as discussed in
that paper.  

(3)~{\it AW~Per}: The Cepheid and its companion have been
rediscussed by Massa \& Evans (2008), who derived a temperature for the hottest
companion of  $T_{\rm eff} = 15735 \pm 248$~K\null.  This effective temperature
is used with the Harmanec relation to derive the mass and spectral type given in
Table~1. Massa \& Evans confirm, however, that the secondary is itself a binary
based on the mass function of the spectroscopic orbit.  

(4)~{\it T~Mon}: From
\HST\/ high-resolution UV spectra, Evans et al.\ (1999) found  that the
companion is a magnetic chemically peculiar Ap star, very similar to
$\alpha^2$~CVn, and also a binary.  The companion mass is taken from that paper.

Table 1 contains a sample of intermediate-mass companions of $\sim$$5\,M_\odot$
stars with uniquely complete information over the full range range of
separations. However, there are some further points that need to be addressed. 
As is typical of massive stars, there is a high fraction of triple  systems in
this list of binaries (Evans et al.\ 2005).  The spectral types and masses in
Table~1 pertain to the hottest companion star in the system, but there may be
additional system members. Two examples are as follows. (1)~W~Sgr is a
spectroscopic binary with a period of 1780~days, and \IUE\/ revealed an A0~V
companion. However, spatially resolved UV spectra obtained by Evans et al.\
(2009) with the \HST\/ Space Telescope Imaging Spectrograph (STIS) showed that
the A0~V star is resolved from the Cepheid at a separation of $0\farcs1645$,
based on an analysis of STIS spectra taken at several telescope roll angles.
Thus the A0~V star is not the secondary component in the 1780-day binary, and
the system is a triple. (2)~V1334 Cyg is a single-lined spectroscopic binary
with an orbital period of  1938~days (Evans 2000), and a B7~V companion detected
by \IUE\null.   V1334~Cyg is cataloged as the resolved double star ADS~14859, 
with several reports of a companion being seen by visual observers at
separations of $0\farcs1$--$0\farcs2$; if so, V1334~Cyg would also be a triple
system.   However, neither \HST\/ FGS interferometry nor FOC imaging 
(1998--2000) were able to resolve the visual  companion, nor were there any
convincing detections of a companion in speckle-interferometry measurements
between 1976 and 2005 (Evans et al.\ 2006). But very recently, Gallenne et al.\
(2013), using the CHARA array, reported that they have resolved a very close
companion in observations made at two epochs in 2012. The measured separations
were $0\farcs00891$ and $0\farcs00836$.  These observations indicate that the B
star seen by \IUE\/ is the 1938-day companion, but this leaves the occasional
reports of a more distant visually resolved companion unexplained.

\section{Orbital Separations}

Having assembled the complete sample of 18 Cepheids brighter than $\langle
V\rangle = 8$ that have binary companions more massive than $2\,M_\odot$, we
will now investigate the orbital separations in these systems.   In col.~5 of
Table~1, we indicate whether the systems have a spectroscopic RV orbit with a
known period (o), have been spatially resolved (r), or have an unknown orbital
period but detected orbital motion (om).

\subsection{Cepheids with Known Spectroscopic Orbits}

Of the 18 systems listed in Table~1, nine have known orbital  periods based on
RV studies. For these binaries, the orbital periods are listed in col.~6 of
Table~1, and are taken from Evans et al.\ (2005) or Evans et al.\ (2011b). The
logarithms of the orbital periods are given in col.~7, and the logarithms of the
orbital separations in col.~8.  These objects tend of course to be the more
compact binary systems.

\subsection{Cepheids in Resolved Binaries}

\subsubsection{HST WFC3 Imaging}

We have recently completed a snapshot imaging survey of 69 nearby Cepheids with
the \HST\/ WFC3 camera (program ID number GO-12215). Full details of the survey,
in  particular point-spread function (PSF) subtraction to search for resolved
low-mass  companions of the Cepheids, will be presented in a later paper.
However, some of the results are relevant to the present study of more massive
Cepheid companions.

The WFC3 images were obtained in the medium-width F621M and F845M filters,
hereafter referred to as ``$V$'' and ``$I$\null.''  All 18 stars listed in
Table~1 were imaged in the course of the snapshot survey.  For three of the
targets---$\eta$~Aql, V659~Cen, and S~Nor---the intermediate-mass companion
stars were resolved. Figure~1 depicts the $V$-band images of these three
systems, and Table~2 gives details of the observations and
measurements.  The companions are
 plainly visible although the PSF is  complicated and even these
 relatively bright companions are significantly fainter than the Cepheid. 
 We did not attempt to  measure the brightnesses of these companions in these
 images, since the \IUE\/ spectra provide information about  the temperature and
 brightness of the companions.  However, we have measured  the separation
 from the Cepheid directly on the $I$-band images,  which is listed in Table~2. 
 These are, of course, only the instantaneous projected separations;
 however we will be examining the  distribution of the logarithm of
 separations, so this is a small uncertainty.

For
completeness, we also include in Table~2 the wider resolved $\delta$~Cep system,
and the close W~Sgr, both of which were discussed in \S2. (The companion of
$\delta$~Cep was outside our WFC3 field of view, and the companion of W~Sgr was
within the saturated pixels close to the Cepheid.)

For $\eta$ Aql, Benedict et al.\ (2007) found perturbations in their \HST\/ FGS
measurements within a couple of years, implying a companion in a relatively
short-period orbit.  Since we have now directly resolved the hottest companion
in the system (Table~2) with a much larger separation/period, we conclude that
the system is triple. The Cepheid S~Nor is a member of a cluster, increasing the
probability of a chance optical alignment. However, the small separation
($0\farcs9$) makes a physical association highly probable. S~Nor also has a hot
companion at a much wider separation of $36''$ (Evans \& Udalski 1994), making
it a possible triple. In this case, given the high stellar density in the cluster,
this could be a chance alignment.
 
\subsubsection{Approximate Orbital Periods}

We used the angular separations in Table~2, along with the secondary masses from
Table~1, and the primary masses and distances from Table~3 (below), to calculate
nominal orbital periods, by equating the projected angular separation to the
semimajor axis, $a$, of the putative orbit.  The resulting $\log P$ values are
given in col.~7 of Table 1. To distinguish these from the directly determined
spectroscopic periods (\S3.1), the $\log P$ values are given to only one decimal
place. Col.~8 of Table~1 gives the values of $\log a$, again to only one
decimal.

\subsection{Cepheids with and without Detected Orbital Motion}

Of the 18 systems in Table~1, nine have known orbital periods, and five have
been spatially resolved, as recounted above. The remaining four stars (SU~Cas,
BP~Cir, T~Mon, and T~Vul)  have detected hot companions whose temperatures and
luminosities are consistent with the distances of the Cepheids (Evans 1992b;
Evans 1992c; Evans 1994; Evans et al.\ 1999). While a chance alignment between
a B or A star and a Cepheid is highly improbable, orbital motion would be
conclusive proof of physical association. In this subsection we discuss what is
known from RVs in the literature, and what limits can be put on the
separations.  

(1)~{\it SU Cas}:  RVs have been measured in a number of studies.  The best
 claim for the detection of orbital motion is by Gorynya, Rastorguev,
\& Samus (1996), who rate SU~Cas as a possible spectroscopic binary.  We have
tested this  by comparing two seasons of accurate  data from the same group
(Moscow University) so that instrumental differences should  be minimal. Typical
uncertainties of their annual velocities are $\pm$$1\,\rm km\, s^{-1}$. We  have
chosen data from two years (1995 and 1997) which are  predicted to have orbital
velocities close to minimum and maximum according to their proposed orbit, and a
velocity difference of $6\,\,\rm km\, s^{-1} $.   When the pulsation velocity
curves from the two years are overlaid, there is no appreciable difference,
certainly nothing as  large as that predicted by the orbit.  We conclude that
orbital motion has {\it not\/} been detected convincingly.

(2)~{\it BP Cir}: RV data have been  discussed by Petterson et al.\ (2004). 
The  original velocity data are from Balona (1981), and have standard deviations
of  $2.5\,\rm km\, s^{-1}$ (Stobie  \& Balona 1979).  Data were added from the
Mount John University Observatory, with the final three years providing an
accuracy of $\pm$$0.3\,\rm km\, s^{-1}$. Petterson et al.\ estimate the orbital
motion to be greater than $5\,\rm km\, s^{-1}$.   Orbital motion appears to be 
seen on timescales of decades, providing some constraint on the period, but
further observations are needed for confirmation. 

(3)~{\it T Mon}: RV variation is  seen, although the orbital period is too long
for a determination at present. Preliminary estimates of the period are between
90 and 260 years (Evans et al.\ 1999).  

(4)~{\it T Vul}:  It has been suggested several times that orbital motion may
have been  detected in RV measurements. Bersier et al.\ (1994) discuss this on
the basis of 11 years of CORAVEL data.  They find a standard deviation of the
data around the pulsation Fourier   curve of only $0.55\,\rm km\,
s^{-1}$.  There are some
limitations in the spacing of the data,  in that the CORAVEL observations were
made only in the autumn and the most likely suggested  period is close
to  2~years. 
However, there is no evidence of orbital motion at the level of $<$$1\,\rm
km\, s^{-1}$.   Kiss (1998) and
Kiss \& Vinko (2000) extended the data series to 20 years using RVs from David
Dunlap Observatory spectra.  Fig.~3 in Kiss \& Vinko shows no indication of 
orbital motion, only a possible small difference in the shape of the curve from
the analytic fit to the CORAVEL data.  We conclude that the highest-quality data
show  no orbital motion. 

\subsubsection{Approximate Orbital Periods}

While these four stars do not have a period or separation as well  defined as
either the stars  with RV orbits or the resolved stars, there is information on
both  these quantities which  provides significant constraints. For the two stars
with orbital motion, BP~Cir and T~Mon,  we can assign reasonable estimates of
the periods from the discussion in the previous section which  should not result
in large errors in the distribution of $\log P$ for the entire sample. For
BP~Cir, an orbital period of 20 years ($\log P = 3.9$ in days) is plausible for
the observed orbital motion.  For T~Mon, a period of $\sim$150~years ($\log P =
4.7$ in days) is in the middle of the range of plausible orbital periods.  

The remaining two stars---SU Cas and T Vul---appear to lie in the orbital
separation range  between the stars with known orbital periods  and those for
which the companions are separated widely enough to be resolved with \HST\/ or
from the ground. Neither star was resolved in our WFC3 survey.  We
estimate that these comparatively bright companions would have been
resolved for a separation of more than $0\farcs3$. 
Using the distances to
the two stars, as well as the masses (Tables 1 and 3), this results in upper
limits to $\log P$ of 5.2 for SU~Cas and 5.5 for T~Vul.   For lower limits to
the separation, we use RV observations.  As discussed  above for SU Cas, orbital
motion was not seen in recent RV data (Gorynya et al.\ 1996). There are also
earlier high-quality RV data.  Based in particular on the discussion  of Niva \&
Schmidt  (1979) we conclude that no orbital motion has been detected over 40
years, and use that as an orbital period lower limit ($\log P =4.2$ in days). 
Thus the available data  constrain the periods for both stars to lie between
$10^4$ and 3 $\times$ $10^5$~days.  In Table~1 we assign  SU~Cas to $\log P = 5.1$ and
T~Vul to $\log P = 4.9$.

\section{Mass Ratios}

The next parameter of the sample to examine is the mass ratio $q = M_2/M_1$,
where $M_2$ is the mass of the secondary companion and $M_1$ is the mass of the
Cepheid.   One of the strengths of the present study is that the masses of {\it
both\/} the primary and the secondary can be inferred from uncontaminated 
spectra and photometry of both stars in the visible and the UV, respectively. 
Furthermore, this  direct access to the parameters of both components is
available at {\it all\/} orbital separations, which is unique among samples of
massive stars. 

Table~	3 lists the  relevant parameters for the Cepheid components, which have
been determined as follows. Col.~2: pulsation periods; three of the stars (SU
Cas, BP Cir, and V1334 Cyg) pulsate in the first overtone, so the listed period
has been ``fundamentalized'' using the relation from Alcock et al.\  (1995):
$$P_1/P_0 = 0.720 -0.027 \log P_0 \, ,$$
where $P_1$ is the first overtone-mode period and $P_0$ is the
fundamental  period.
Col.~3: unreddened visual absolute magnitude, 
$M_V$, derived from the Leavitt
(period-luminosity) relation as given by Benedict et al.\ (2007):
$$M_V = -4.05 -2.43 (\log P -1.0)\, .$$
Cols.~4--6:
values of $E(B-V)$, $(B-V)_0$, and $\langle V\rangle_0$, which have been
corrected for the  effect of the companion and are taken from the same sources
as the companion spectral types (or can be directly traced from those
references). Corrected photometry for S~Mus is from Evans, Massa, \& Teays
(1994).    The exception is $\delta$ Cep, where the companion does not affect
these values because it is well resolved; its parameters have been taken from
the Galactic Cepheid database\footnote{available at {\tt
http://www.astro.utoronto.ca/DDO/research/cepheids}} (Fernie et al.\ 1995).  For
all of the Cepheids, $\langle V\rangle_0$ has been computed using $R =
A_V/E(B-V) = 3.46$, appropriate for Cepheids (Evans 1991). 
Cols.~7--8: bolometric correction, taken from Flower (1996),
and the resulting value of $\log L/L_\odot$. Col.~9: distance, calculated from
$M_V$ and $\langle V\rangle_0$. Col.~10: mass, computed using the models of
Prada Moroni et al.\ (2012) with moderate convective overshoot [0.2 times the
pressure scale height at the edge of the convective core on the main sequence
(``noncanonical''; Bono 2012 private communication)], from the relation
$$\log M/M_\odot = 0.297 \log L/L_\odot -0.259\, .$$
Using the Cepheid masses in the final column of Table~3, we calculated
values of the mass ratio $q$, which are listed in col.~9 of Table~1.

Figure~2  shows the distribution of Cepheid masses (solid red line) and
companion masses (dashed black line) in our sample, which of course is truncated
for companion masses lower than $2\, M_\odot$. Fig.~3 plots the companion masses
against the Cepheid masses, and indicates no strong correlation between them. In
fact, two of the most massive Cepheids (S~Mus and S~Nor, at masses of 6.2 and
$6.3\,M_\odot$ respectively) have companions covering nearly the full range of
companion masses (5.3 and $2.4\, M_\odot$ respectively.)

\section{Distribution of Separations and Orbital Periods} 

Table 1 provides the information for a study of the distribution of orbital
separations and periods. However, we must keep in mind two sample biases:
(1)~Although the photometric approach to creating the sample means that
companions at any separation are equally likely to be identified, the sample
contains only companions hotter than early A spectral type, with the least
massive being $2.1\, M_\odot$ (T~Vul~B)\null. (2)~Stars evolving off the main
sequence in relatively close orbits will undergo Roche-lobe overflow and the
subsequent evolution of the system will be drastically altered. For Cepheids we
have a good estimation of where this effect sets in. Z~Lac---not in our sample
because the mass of its unseen companion has an upper limit of $1.9\,M_\odot$---
is the Cepheid with the shortest known orbital period (382~days; Sugars \& Evans
1996). It is also the only known Pop I classical 
Cepheid binary orbit with zero eccentricity,
suggesting that it was circularized, appropriate for a system which just missed
significant Roche-lobe overflow. 

The orbital separations in col.~8 of Table~1 were calculated as follows.
Directly measured (projected) separations for the five resolved binaries have
been taken from Table~2. For the remaining objects, the semimajor axes were
calculated from the masses and orbital periods. Values of the separation from
measured orbital periods are given to two decimal places, and from projected
separations or the estimated periods of SU~Cas and T~Vul to one decimal place. 
For the triple systems  the total mass of the system, and  hence the separation,
will be underestimated since we do not know the mass of the third star. 
However, since we know the masses of the two most massive stars in the system,
this is a relatively small underestimate.   Fig.~4 shows the distribution of
separations.  The vertical dashed line marks the cutoff in separations for
periods of less than a year.

The distribution of orbital periods for our Cepheids is shown as the red
histogram in Fig.~5.  We know the bin for $\log P = 2$ to 3  is
incomplete due to the destruction of short-period orbits (Sugars \&
Evans 1996).  Because the shortest orbital period for a Pop I
classical 
Cepheid (Z
Lac, log P = 2.58) falls in the center of this bin, 
we have doubled the number of Cepheids in that bin to account for
stars removed from the sample..  For comparison, we show
(dashed black histogram) the distribution of orbital separations for  solar-mass
stars from Raghavan et al.\ (2010).  We have used their Fig.~11 to  create a
sample with $q > 0.4$ for comparison with the Cepheid sample. (Binaries in the
Raghavan sample with $\log P < 2$ have been omitted.)

The difference between the distributions of orbital periods
for $5\, M_\odot$ and $1\, M_\odot$ in
Fig.~5 is striking, in that the more massive stars are concentrated at shorter
periods than are the solar-mass stars.  The difference in the distributions is
confirmed by the  cumulative distribution functions (CDFs) plotted in  Fig.~6,
showing the observed Cepheid distribution  compared with the Gaussian fitted by
Raghavan et al.\ (2010) to solar-mass stars. A Kolmogorov-Smirnov (K-S) 
test confirms
that the Cepheid CDF is significantly different from the CDF created from
Fig.~11 of Raghavan et al.\ for the same range of mass ratios and separations. 
However if $\log P$ is arbitrarily increased for the Cepheids, the form of the
CDF  closely matches that for the solar-mass stars.  
Sana \& Evans (2011) have assembled binary/multiple properties of O
and early B (down to B3) stars from the field and a large sample of
galactic clusters.  Fig.~7 shows the Cepheid CDF
compared with  the Sana \& Evans (2011) results.  They
fit the data  with a ``broken \"Opik's law''  divided for periods
longer and shorter than 10 days. (\"Opik's law [\"Opik 1924], as
discussed  by Sana
\& Evans, models the distribution of periods to be flat in $\log P$ space.)
Fig.~7 shows the slope of their CDF (defined
for $\log P = 1.0$ to 3.5, but shown in the plot from $\log P= 2.0$ to 3.5,
and  extrapolated to longer periods).  The Cepheid data
fit the extrapolation well to about $\log P = 6.0$.  The decline at higher periods
matches the expectation from  the Gaussian fit to solar mass
stars. For $\log P < 2.5$  Cepheid binaries have low frequency because
Roche-lobe  overflow occurs during post-main-sequence evolution.

One parameter that is similar between  5 and $1\, M_\odot$ stars is
the  binary frequency.  The binary frequencies for
periods longer than a year and $q > 0.4$ are 24\% for Cepheids and 27\% for
solar-mass stars.

\section{Distribution of Mass Ratios} 

The mass ratios, $q$, in Table~1 have been computed directly from {\it
uncontaminated\/} information about  both components.  The mass ratio of BP~Cir
has the unphysical result that it  is slightly larger than 1.0 ($q = 1.05$).  We
take this simply to indicate an uncertainty in the derivation of the masses, and
that the mass of the Cepheid is only very slightly larger than that of
the companion.  
Fig.~8 shows the frequency distribution of $q$.  Although the sample is modest
in  size, it is well defined.   The $q =  0.3$--0.4 bin is presumably somewhat
incomplete, since the sample  criterion was based on the mass of the secondary,
and the resulting $q$ depends also on the  mass of the primary. The highest
frequency is for systems in the smallest two bins ($q = 0.3$--0.5).   One
interpretation of the Cepheid  distribution would be a bimodal one, with a
concentration at large $q$ (equal masses) and another one around $q\simeq0.4$,
with  fewer systems in between.

Does the distribution of mass ratios depend on the separations of the systems? 
Fig.~8  compares the total sample with the sample of closer systems ($\log P <
4$; scaled by a factor of 2).  The largest change is in the smaller-$q$ bins. 
That is, there is an indication that closer  systems are more likely to have
larger $q$ than wider systems (remembering that the Cepheid  sample is limited
to systems of a year or longer, i.e., $\log P > 2.6$.)

Cepheids (Evans et al.\ 2005) and also solar-mass stars (Raghavan et al.\ 2010)
have a high proportion of triple systems among the multiple systems.
The sample discussed here is a particularly good one for examining
the effects of triple
systems, since information is available about each secondary mass and
 frequently the velocity of the secondary, which  
increases the probability of identifying triple systems.  Cols.~10 and
11 in  Table~1
indicate which of the Cepheid sample are known to be  triple
systems, and the source of this information.
(V1334~Cyg is not classified as a triple as discussed above.)  There
may be additional unrecognized  triple systems in the sample since
complete detection requires extensive observation of  both stars in a binary. 
Fig.~9 shows the distribution of mass ratios for the known members of  triple
systems compared with  the full sample. The large-$q$ systems do {\it not\/}
appear in the sample of  triples.  That is, there is an indication that in a
triple system, the mass ratio between the most massive and the second most
massive star in the system is not as large as in a  simple binary system,
presumably frequently somewhat compensated for by the mass of the third star in
the system.    

The comparison between the cumulative distributions is shown in Fig.~10.  As in
Figs.~8 and 9, differences are suggestive but not statistically
significant in a K-S test.

How does the distribution of mass ratios for our Cepheids compare with those for
other stellar classes?  Sana \& Evans (2011) find that for their O-star sample,
a uniform distribution in $q$ fits the data well.  This is shown in Fig.~11,
compared with the  distribution for the Cepheid sample.  The normalization is
approximate  because the Cepheid study does not include small-$q$ systems.   For
solar-mass stars, Duquennoy \& Mayor (1991) concluded that the $q$ distribution
is very similar to the initial mass function (IMF), with few systems near $q=1$
and many with small $q$.  The recent, updated sample by Raghavan et al.\ (2010)
came to different conclusions.  Fig.~11 also shows the distribution of $q$
values for the Raghavan et al.\ sample of solar-mass stars.  For the comparison
we have used data from their Fig.~16 (left) binary systems.  Their study
illustrates very well the complexity presented by components from the
combination of  binary and multiple systems.  They have looked at the $q$ 
distribution for binary and multiple systems divided in two ways (within a
spectroscopic binary and also the ratio of the combined spectroscopic binary
mass to the mass of the visual binary secondary). The first approach (their
Fig.~16b) gives a distribution very similar to that of their binary systems.  We
do not have information comparable to the second approach (their Fig.~16c). 
Therefore we have adopted their distribution for binary systems as
representative of their findings.  We have rebinned it into bins covering 0.1 in
$q$ and approximately scaled it (Fig.~11).  This treatment of the data decreases
the prominence of the peak at equal masses ($q = 1$) for the solar-mass stars,
which makes the distribution reasonably similar to the uniform distribution for
the O stars.  The largest peak in the Cepheid data is at the smallest $q$
values  in our sample.  Further discussion of the distribution of $q$ values is
deferred to a later paper in this series, which will deal with a larger range of
mass ratios.   

An important parameter to investigate further with this sample is whether there
is  any difference in binary/multiple characteristics between close and wide
systems. Figs.~12 and 13 explore further the comparison made in Fig.~8.  Fig.~12
shows the  distributions of mass ratios for stars with $\log P > 4$ and 
$\log P < 4$.  The suggestion  that closer systems have larger mass ratios is confirmed in
Fig.~13, which compares the cumulative distributions for these two period
groups.

An important question in interpreting the distributions in Figs.~12 and 13 is
whether they trace back to the original formation state, or whether there has
been any dynamical evolution in the periods and separations of the systems after
formation.  One approach to  this problem is to compare the properties of binary
and triple systems.  Systems with three (or more) components can have
interactions between the stars which will  alter the original configuration
(even sometimes ejecting a component), which does not happen in purely binary
systems.  As discussed above, we have identified a number of triple systems
within the sample.  Fig.~14 compares the distribution of mass ratio and $\log P$
for binary and triple systems.  As noted above, the binary systems have higher
$q$ on  average.  Using the binary sample as ``dynamically unevolved,'' Fig.~14
shows that it  contains wide systems as well as close ones, similar to the
triple systems.  That is, the indication is that the wide systems are not
exclusively those with a component  moved out by dynamical interaction.  There
are two caveats to this first exploration.  Some of the binaries may have
undetected third components (despite the extensive  multi-wavelength information
on the sample).  However, the binary sample should be  dominated by systems
which have not had internal interactions between components. Second, of course,
it is possible that a previous component has been ejected, but  this could have
happened in either the binary or triple sample.

\section{Summary} 

We have used an \IUE\/ survey of Cepheids to create a list of binary systems
with mass ratios $q \ge 0.4$, which is complete for all separations. We have
combined separations from resolved companions from an \HST\/ imaging survey with
orbits from the literature and RV data to derive the distribution of orbital
separations for the sample. The $5\, M_\odot$ Cepheids are found to have
systematically shorter orbital periods than
the sample of $1\,M_\odot$ stars from Raghavan et al.\ (2010),
confirmed as statistically significant by a K-S test (Fig. 5). The distribution
of mass ratios is also presented, with suggestions that closer systems
have larger mass ratios and also that triple systems have smaller
secondary to primary mass ratios.  The distribution of mass ratios as a
function of orbital separation, however, is the same whether a system is a
binary or a triple.  

These results for $5\, M_\odot$ for all separations and mass ratios $q
\ge 0.4$ is the first step to be followed by studies of resolved
companions, low-mass companions (of late B stars), and radial-velocity
observations. The picture that is thus built up of binary properties
will be compared with those of higher and lower mass stars for
insights into star formation as well as future evolution.








\acknowledgments

It is a pleasure to thank H. Harris and K. Kratter for valuable discussions.
\IUE\/ continues to provide a valuable foundation for Cepheid companion
studies.  
 Support for this work was also provided by HST grant GO-12215.01-A and
 from the Chandra X-ray Center NASA Contract NAS8-03060.  
Vizier and  SIMBAD were used in the preparation of this study.

{\it Facilities:} 
\facility{HST:WFC3}, \facility{IUE}

\clearpage

\begin{figure}
 \includegraphics[width=\columnwidth]{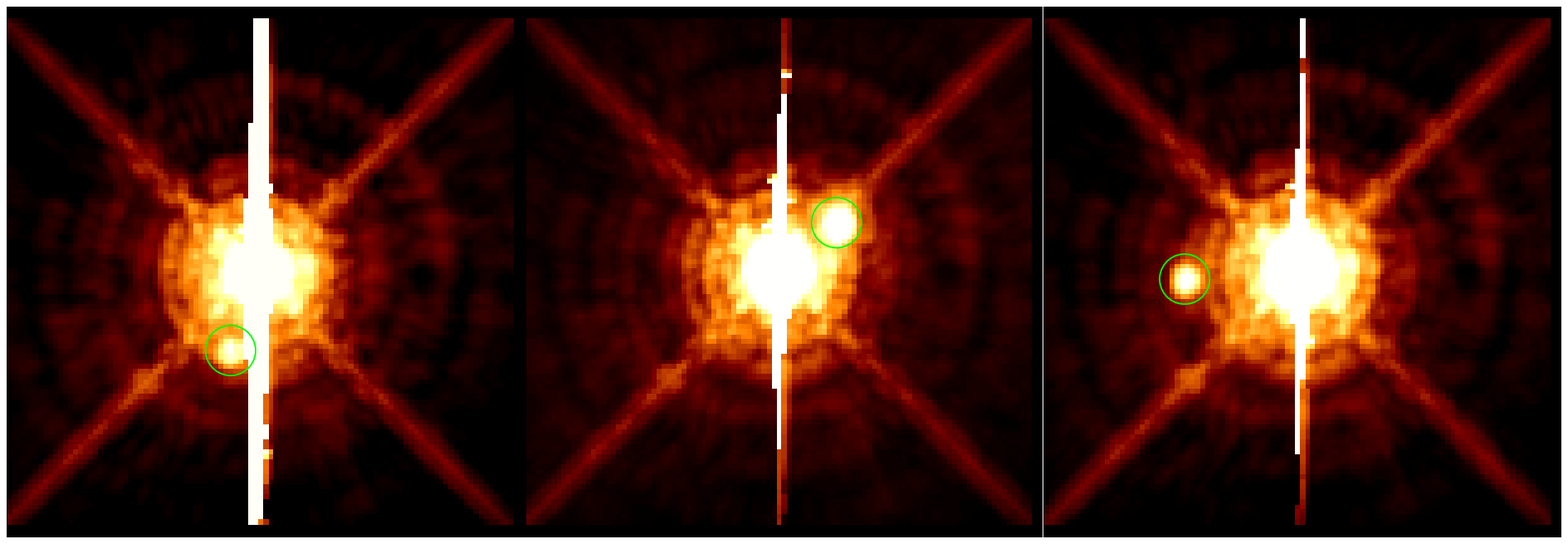}
\caption{\HST\/ images of three Cepheids whose hot companions were
resolved in WFC3 images: $\eta$~Aql
(left), V659~Cen (center), and S~Nor (right).  These are $V$-band images, with a
logarithmic stretch. Each frame is $4'' \times 4''$.  Companions are
    circled in green.  \label{fig1}}
\end{figure}

\begin{figure}
 \includegraphics[width=\columnwidth]{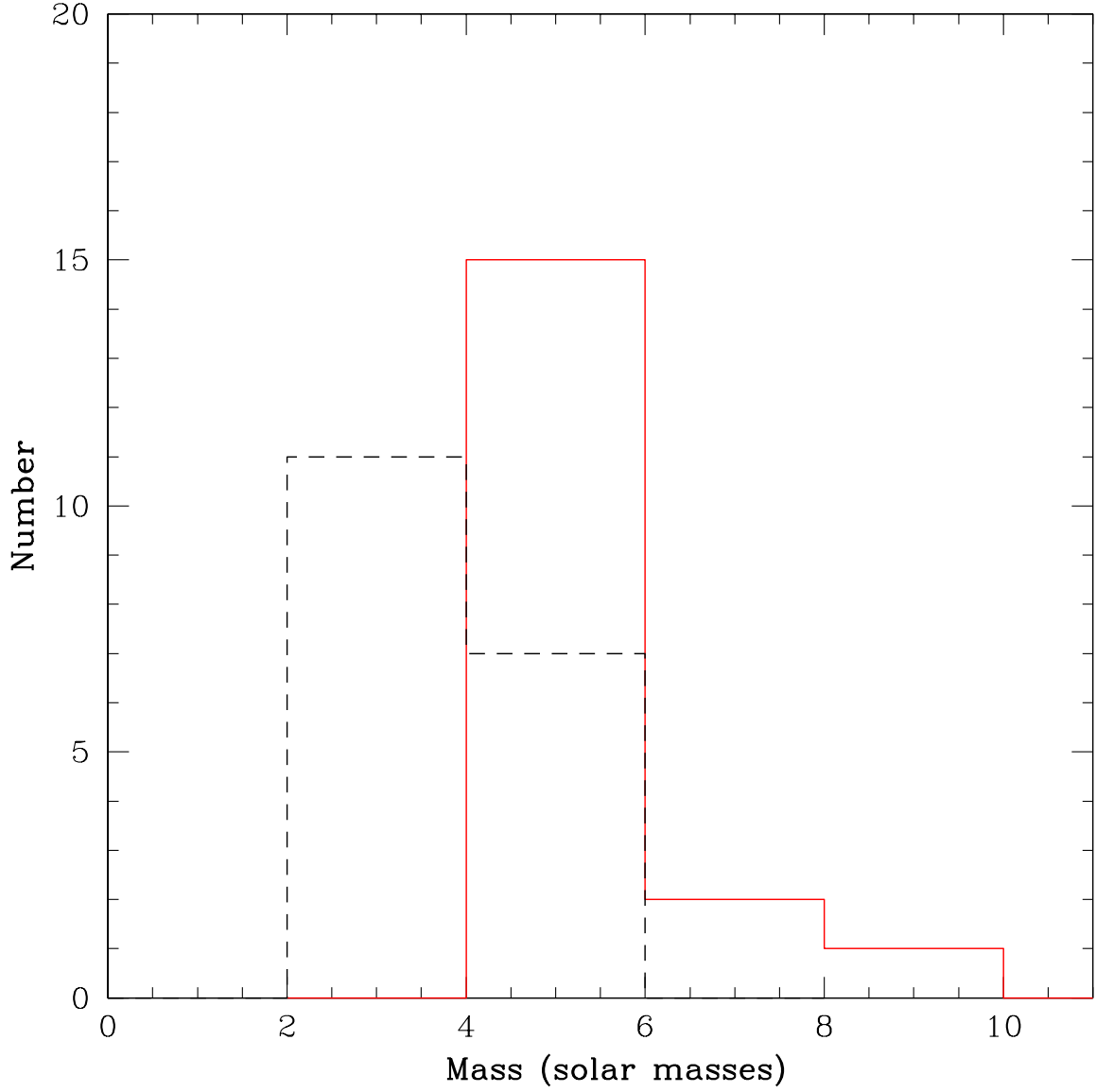}
\caption{The distribution of masses: Cepheid masses: solid line;
  companion masses: dashed line. \label{fig2}}
\end{figure}

\begin{figure}
 \includegraphics[width=\columnwidth]{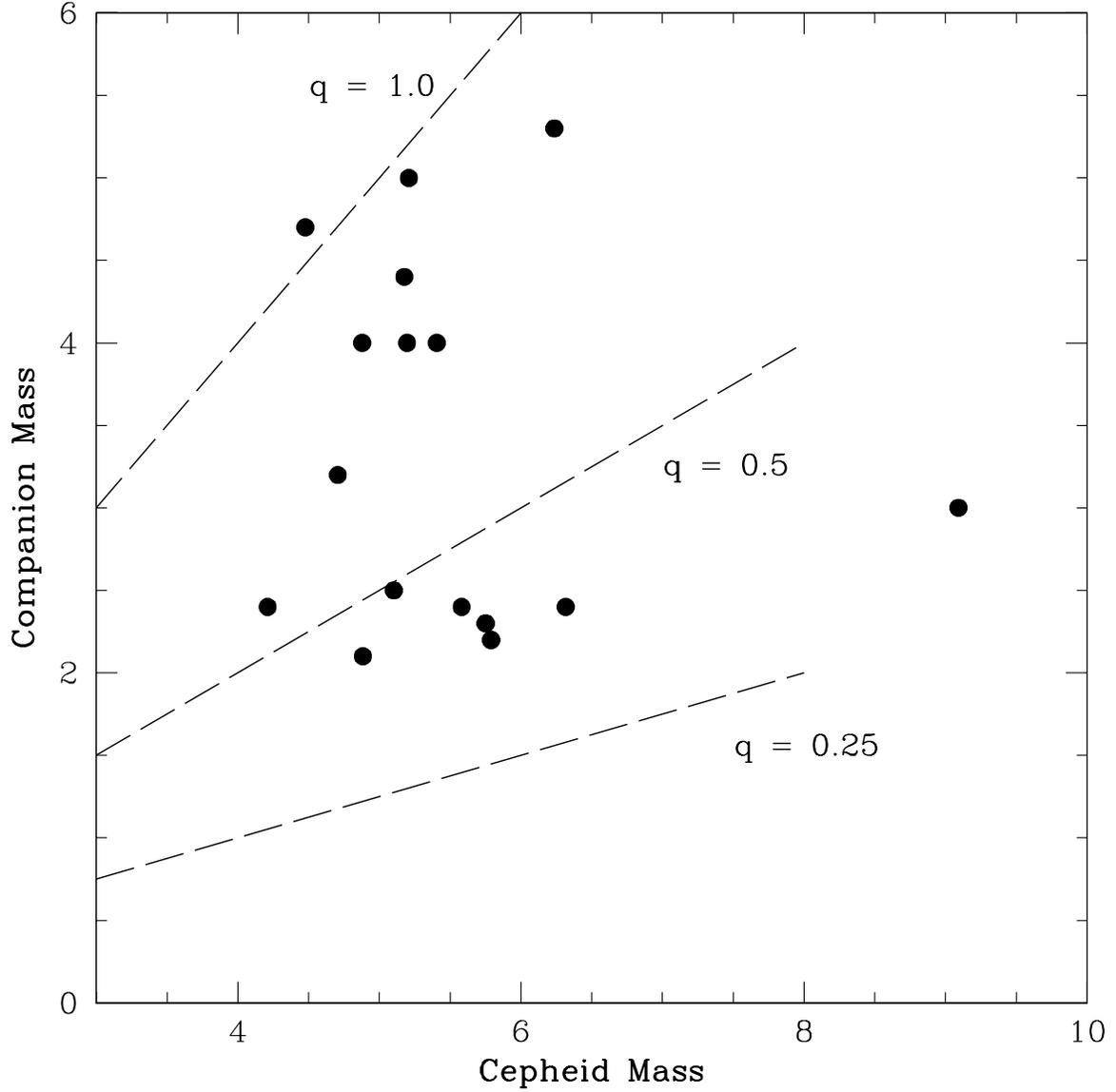}
\caption{Mass of the companion as a function of the mass of the 
Cepheid.  Masses in all figures are in solar masses.  Dashed lines
indicate mass ratios. \label{fig3}}
\end{figure}

\begin{figure}
 \includegraphics[width=\columnwidth]{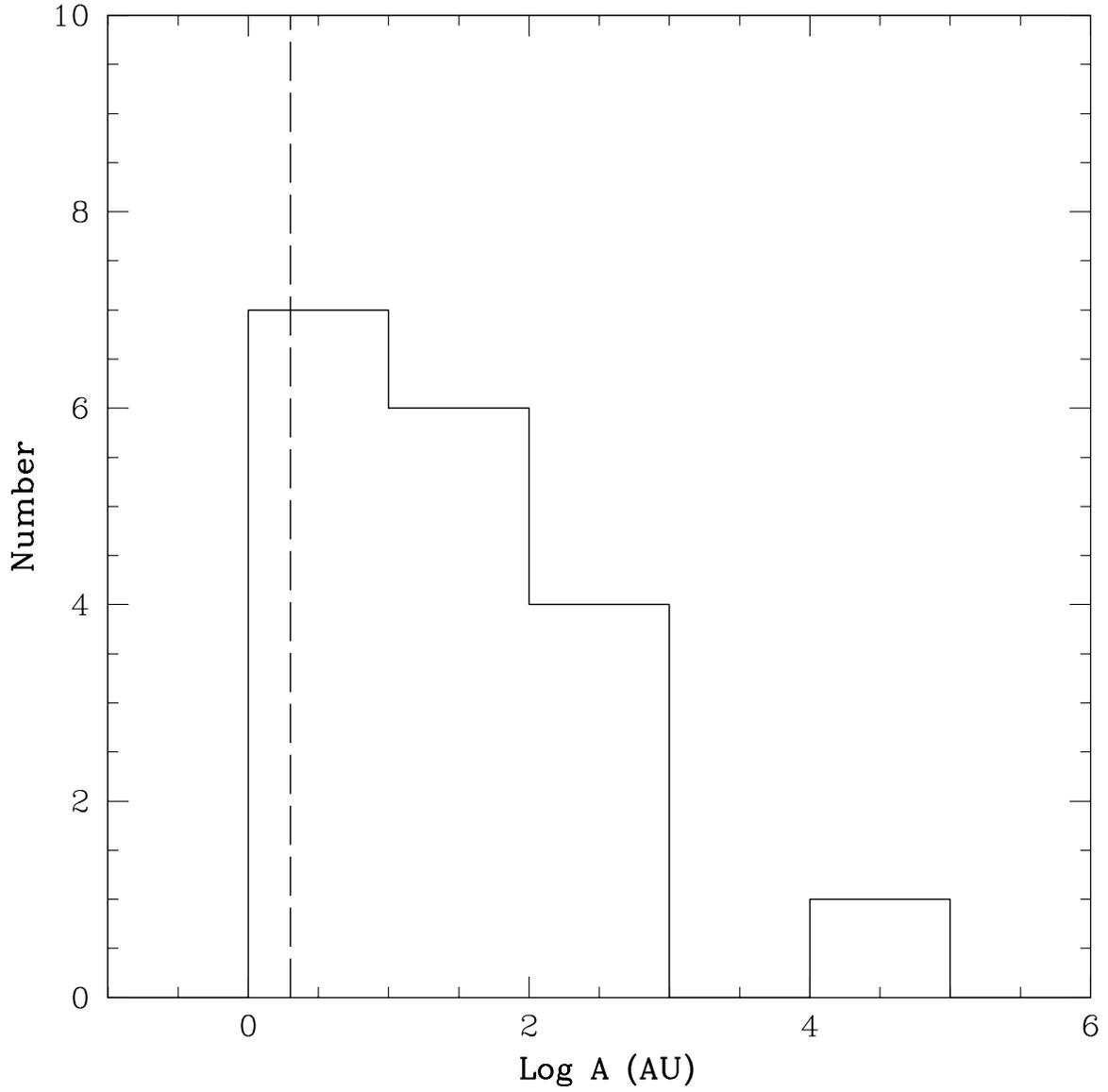}
\caption{The distribution of  separations ($\log a$ in AU)\null.   The dashed
vertical line indicates the separation for periods of 1 year, below which
Cepheid orbits have been disrupted by Roche-lobe overflow.  \label{fig4}}
\end{figure}

\begin{figure}
 \includegraphics[width=\columnwidth]{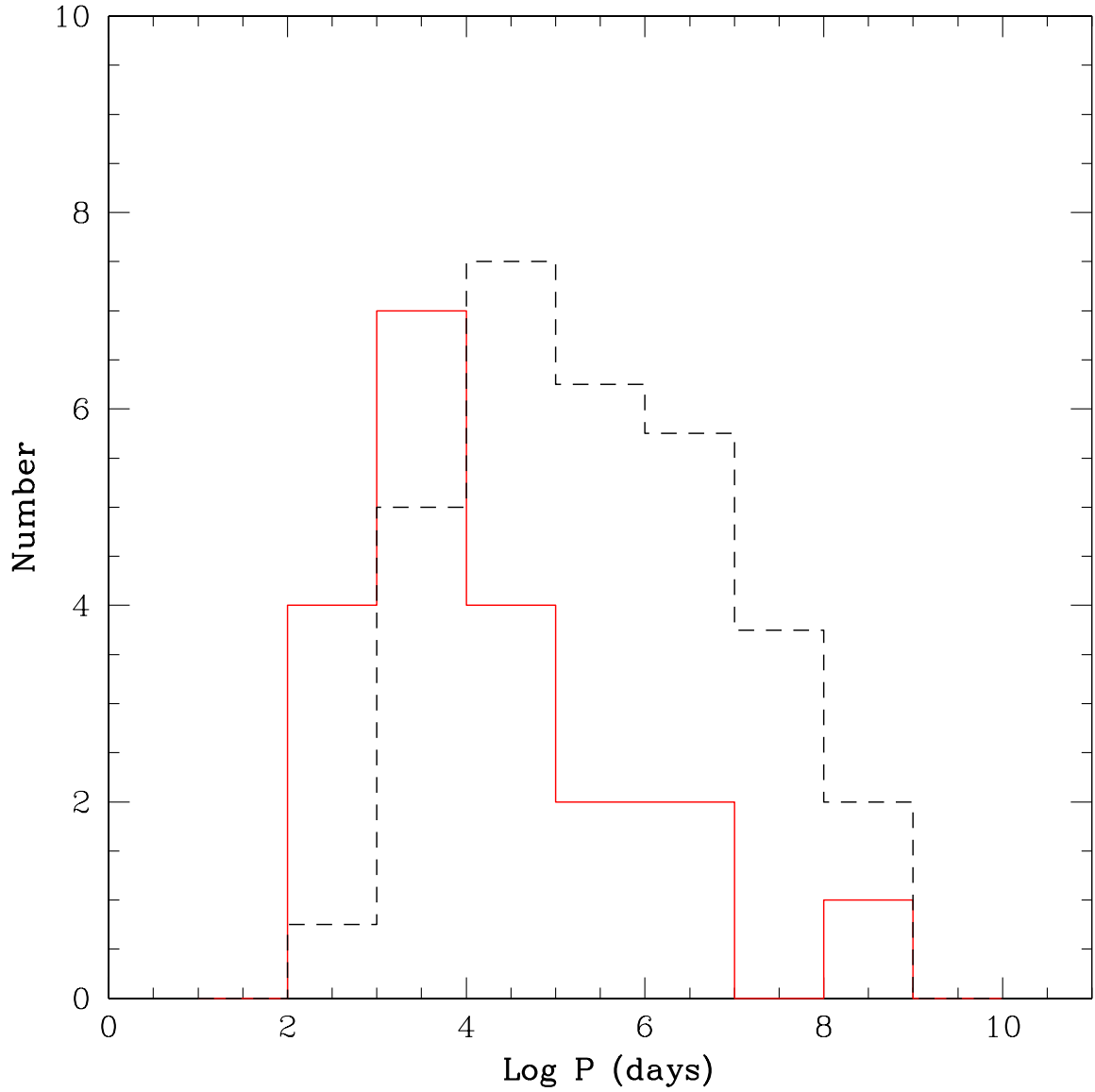}
\caption{The distribution of orbital periods.  Cepheids: solid red line;
 solar mass stars: dashed black line.  Both samples include only systems with 
$q = M_2/M_1$ greater than 0.3-0.4. Binaries in the Raghavan sample
with $\log P < 2$ are not shown. \label{fig5}}
\end{figure}

\begin{figure}
 \includegraphics[width=\columnwidth]{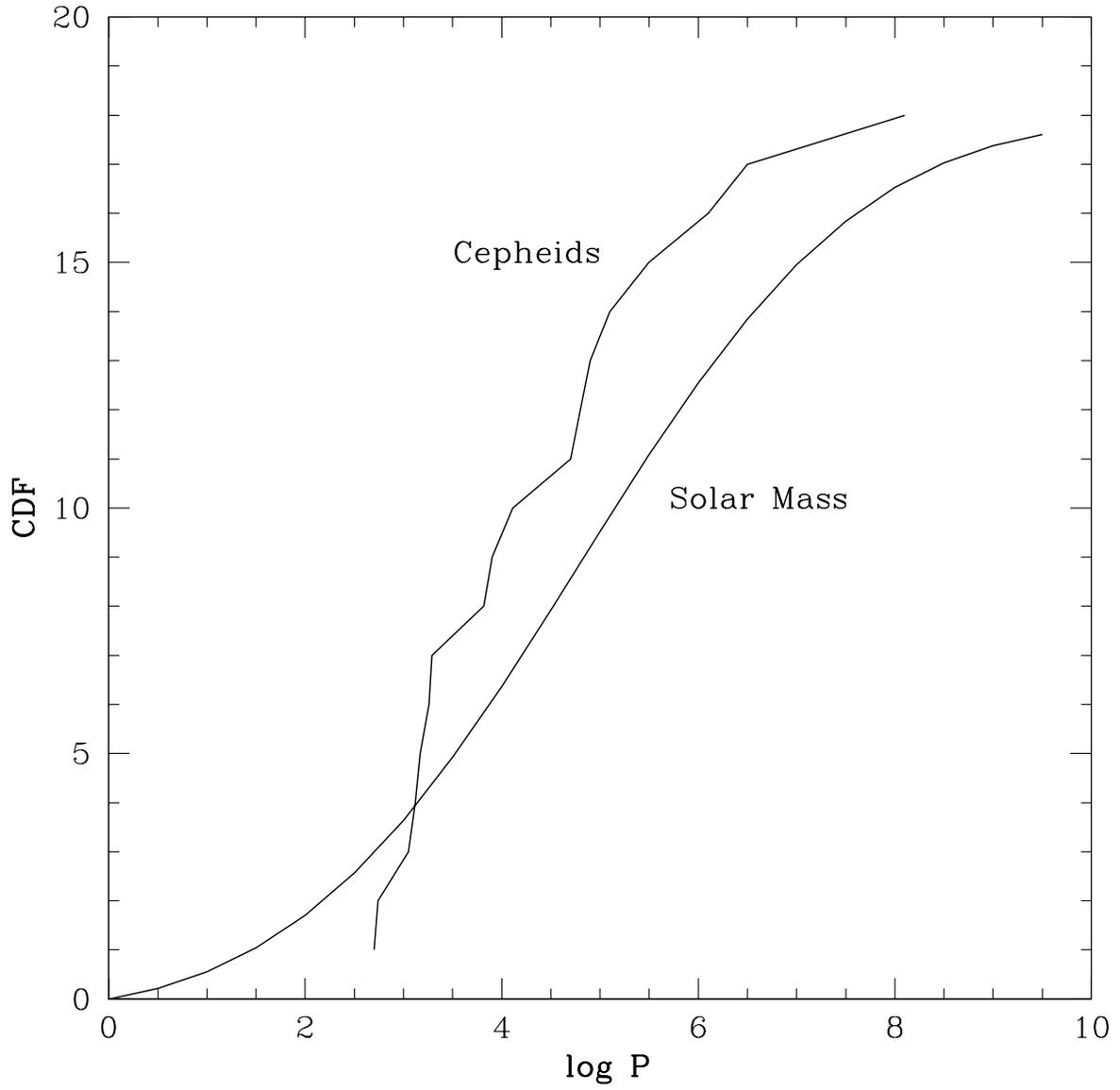}
\caption{The cumulative distribution function (CDF) of
Cepheid orbital periods  and 
solar mass orbital periods as a function of $\log P$ (in days).  \label{fig6}}
\end{figure}

\begin{figure}
 \includegraphics[width=\columnwidth]{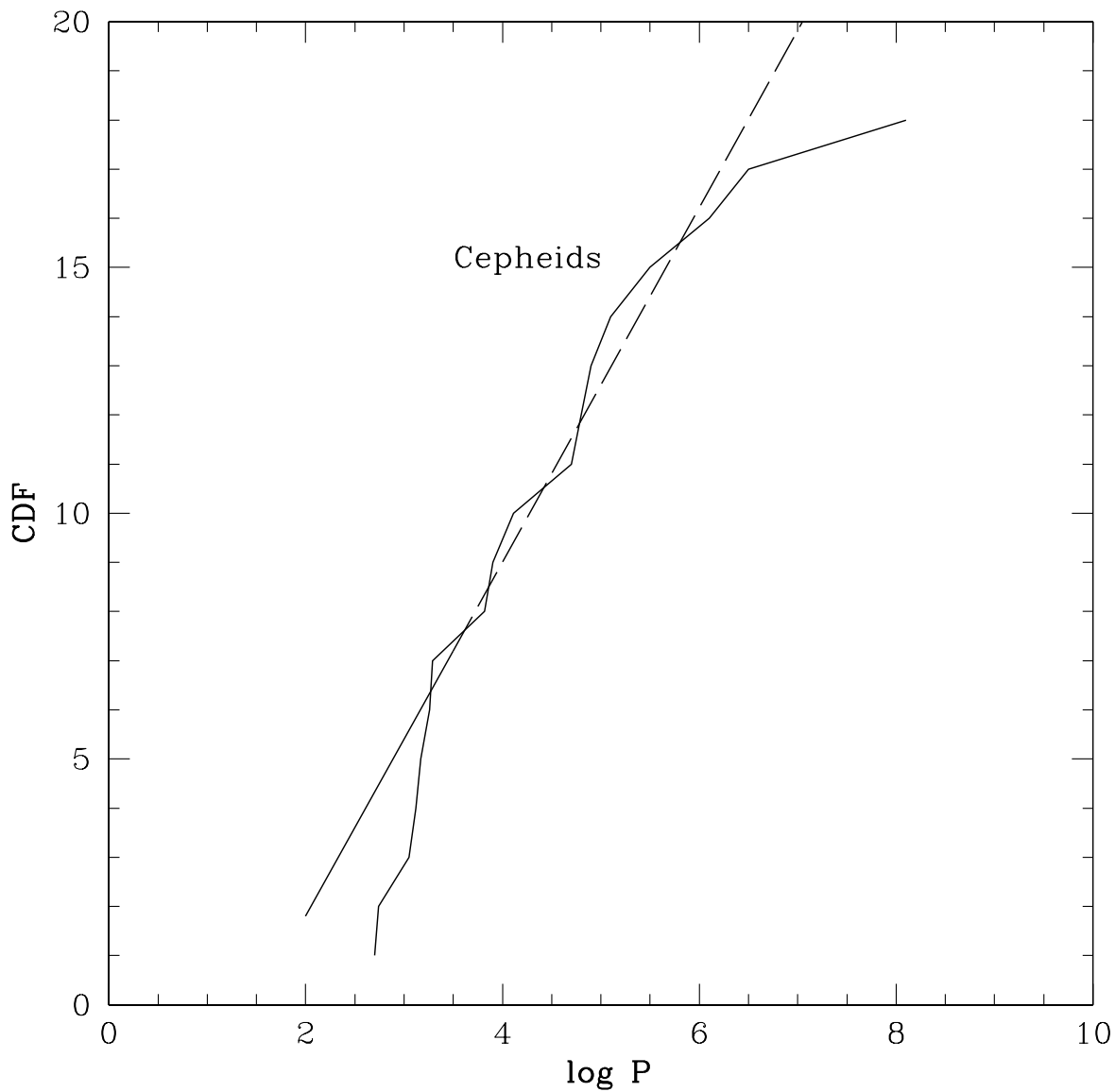}
\caption{ The cumulative distribution function (CDF) of Cepheid periods  compared
with  binaries among massive stars from Sana \& Evans (2011).  
The slope of the broken ``\"Opik law'' for Sana \& Evans  longer periods
is shown by the line.  It  is
solid for the range of periods covered by the Sana \& Evans sample ($\log P = 2$ to 3.5)
and dashed for the extrapolation to longer periods.  \label{fig7}}
\end{figure}


\begin{figure}
 \includegraphics[width=\columnwidth]{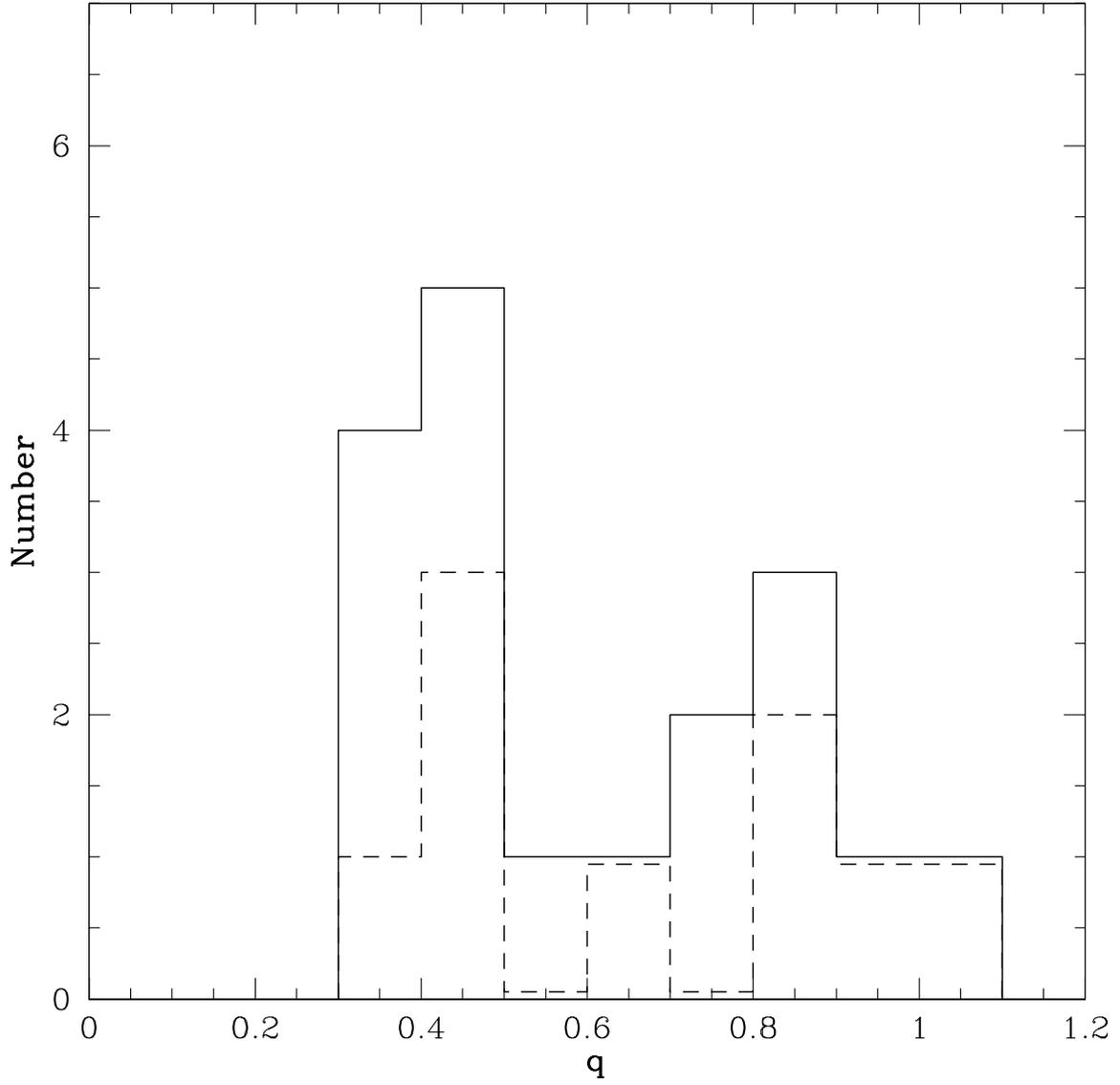}
\caption{The distribution of mass ratios (solid: all stars;
dashed: $\log P < 4$).  Histogram heights have been slightly adjusted 
for clarity.
  \label{fig8}}
\end{figure}

\begin{figure}
 \includegraphics[width=\columnwidth]{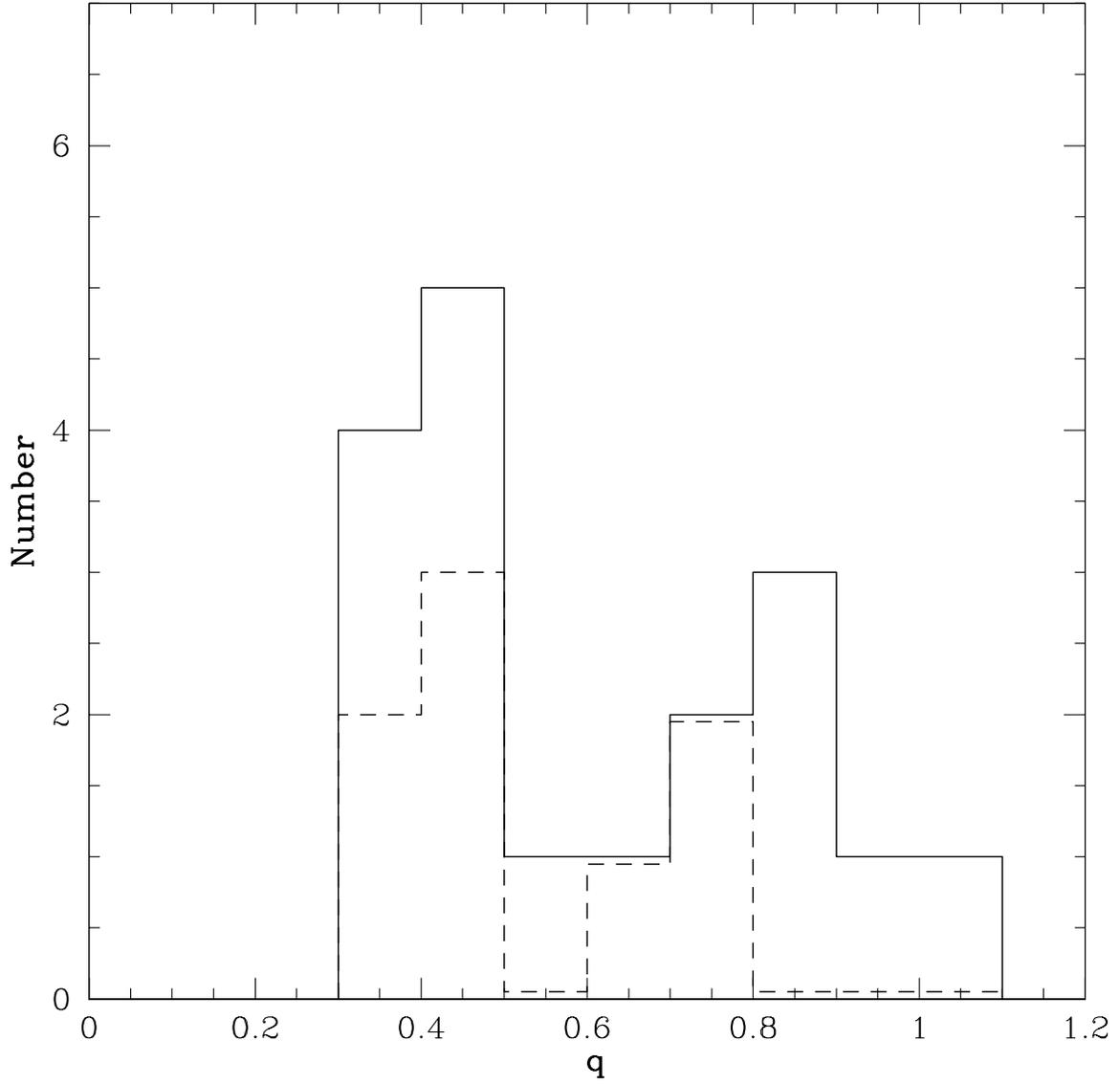}
\caption{The distribution of mass ratios (solid: all stars;
dashed: triple systems).  \label{fig9}}
\end{figure}

\begin{figure}
 \includegraphics[width=\columnwidth]{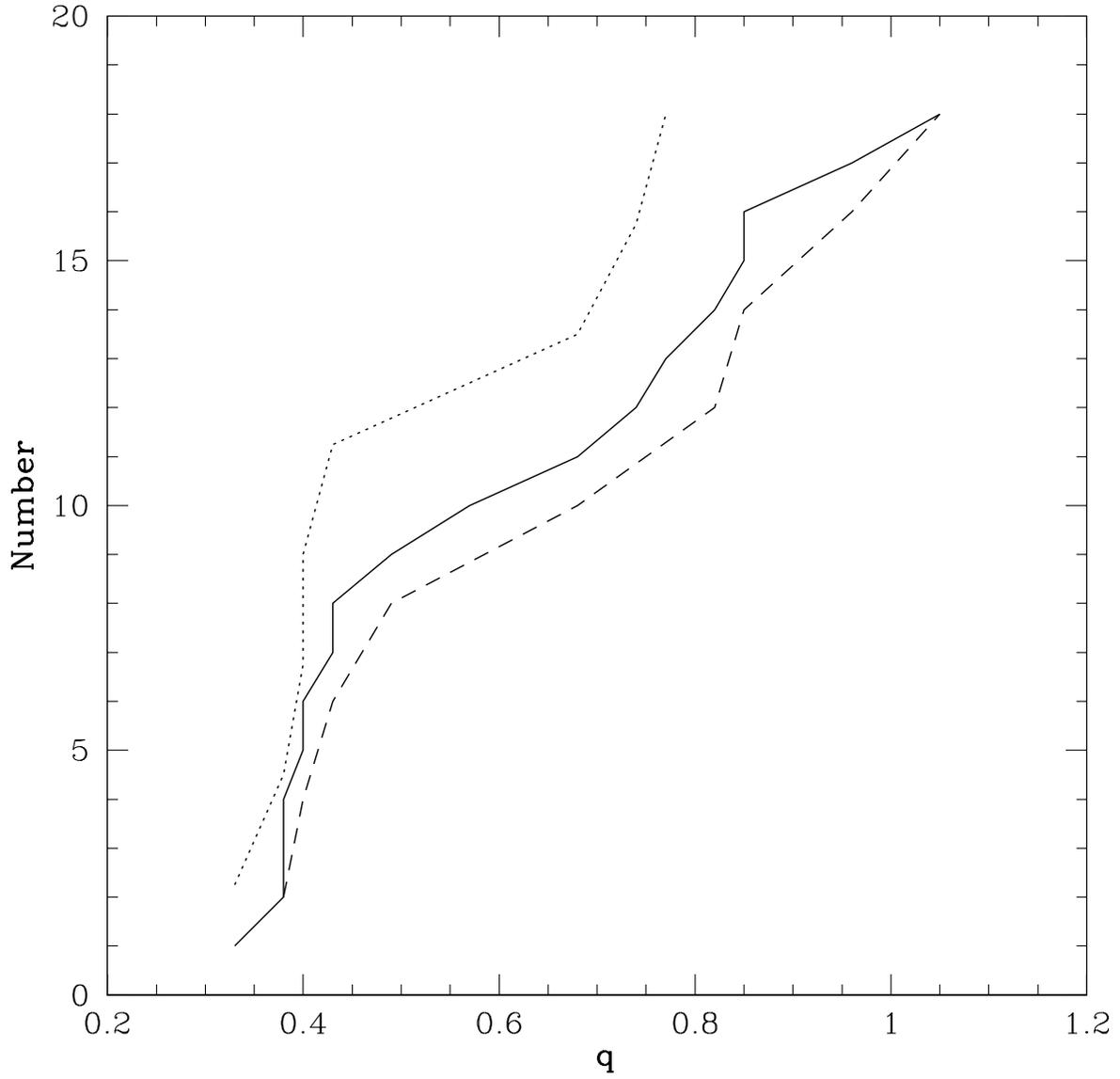}
\caption{Cumulative distribution of mass ratios.  Solid: all stars;
dashed: $\log P < 4$; dotted: triple systems.  The samples of short
period binaries and triple systems have been scaled by 2.0 and 2.25
respectively . \label{fig10}}
\end{figure}

\begin{figure}
 \includegraphics[width=\columnwidth]{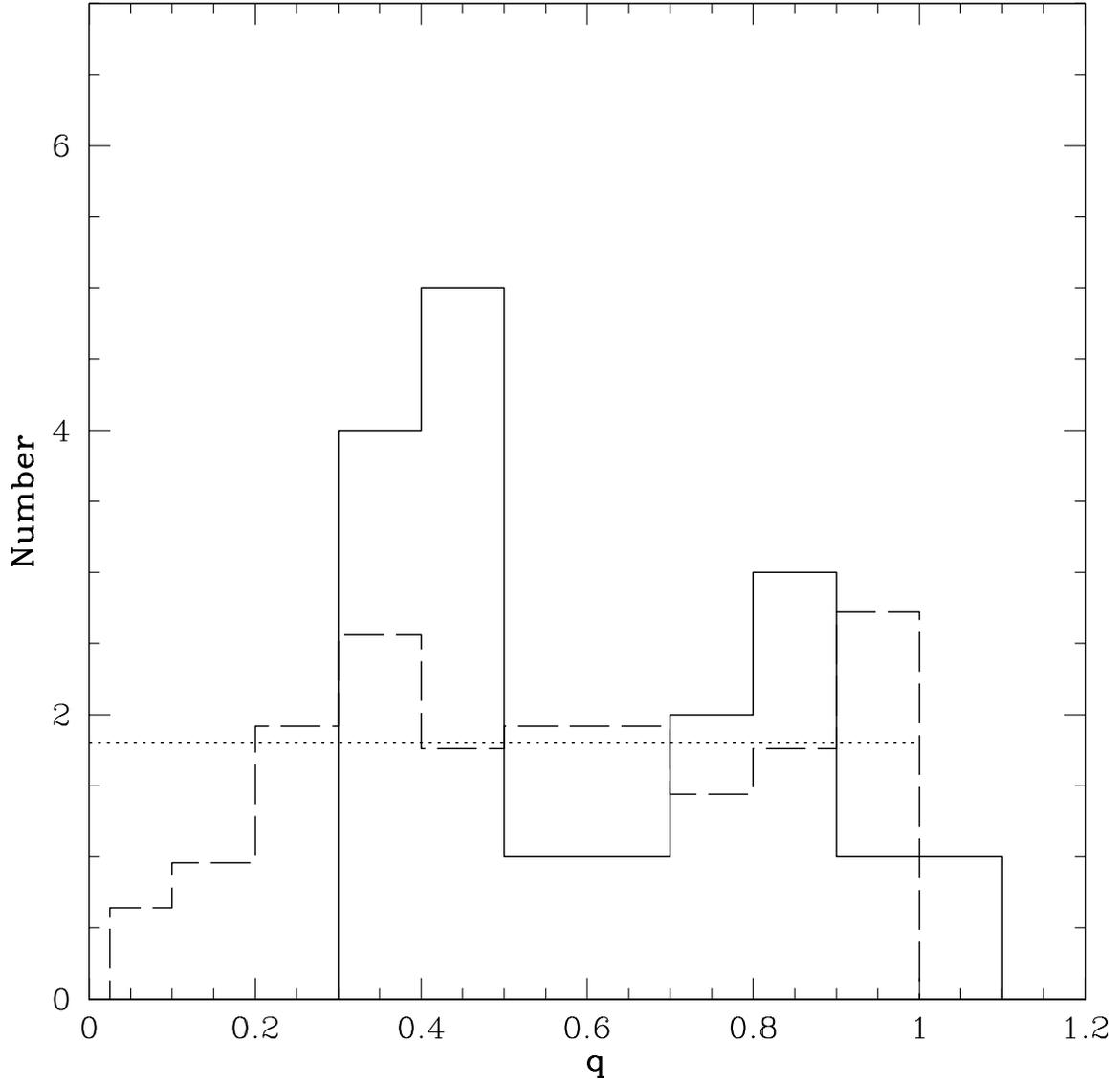}
\caption{The distribution of mass ratios.  Solid line: Cepheids, Fig.~8; dashed
line: Raghavan solar-mass binaries; dotted line: Sana \& Evans O
stars. 
 \label{fig11}}
\end{figure}

\begin{figure}
 \includegraphics[width=\columnwidth]{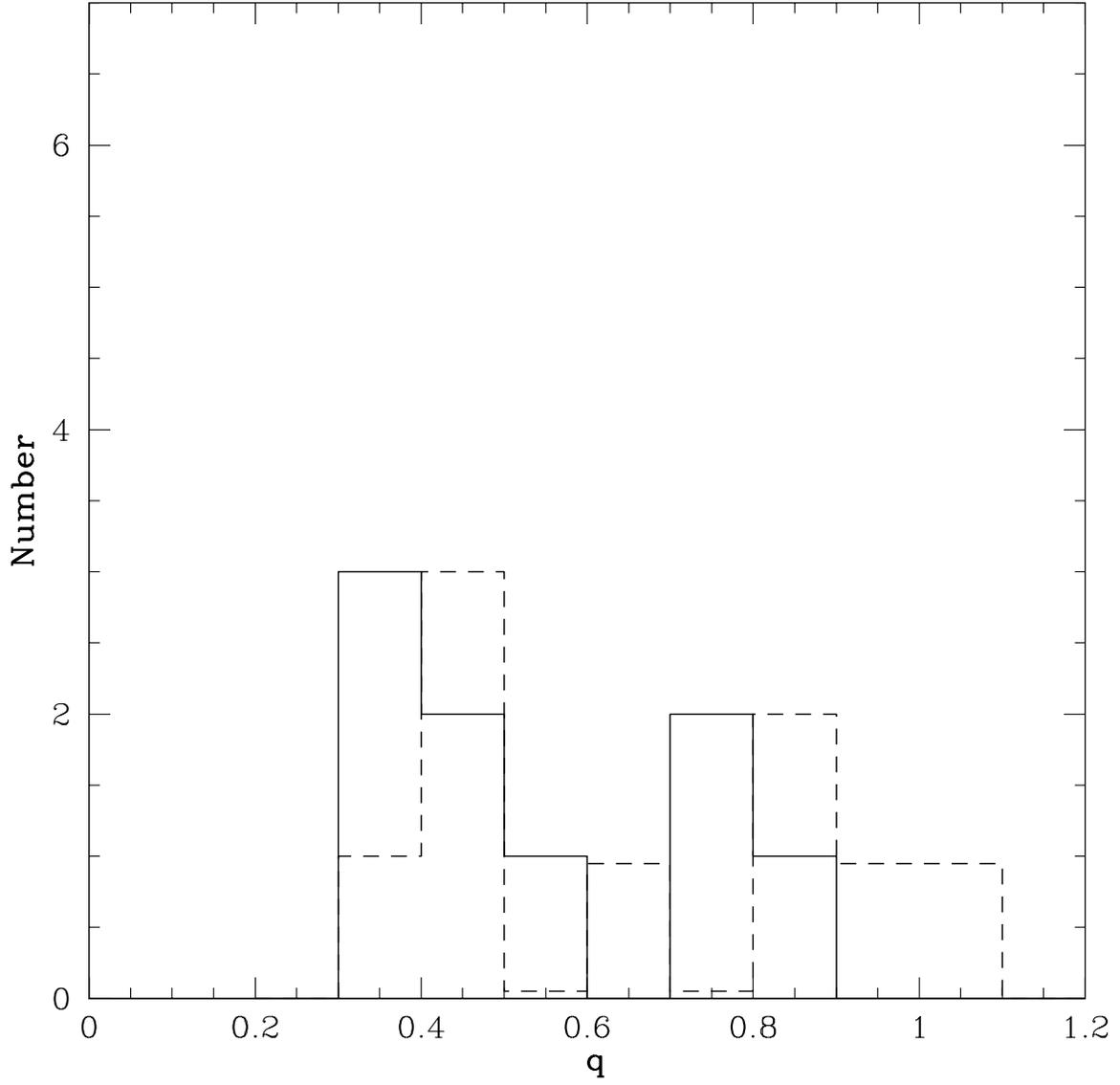}
\caption{The distribution of mass ratios divided into period groups  
($\log P <4$:
dashed line; $\log P >4$: solid line). Periods are in days.
 \label{fig12}}
\end{figure}

\begin{figure}
 \includegraphics[width=\columnwidth]{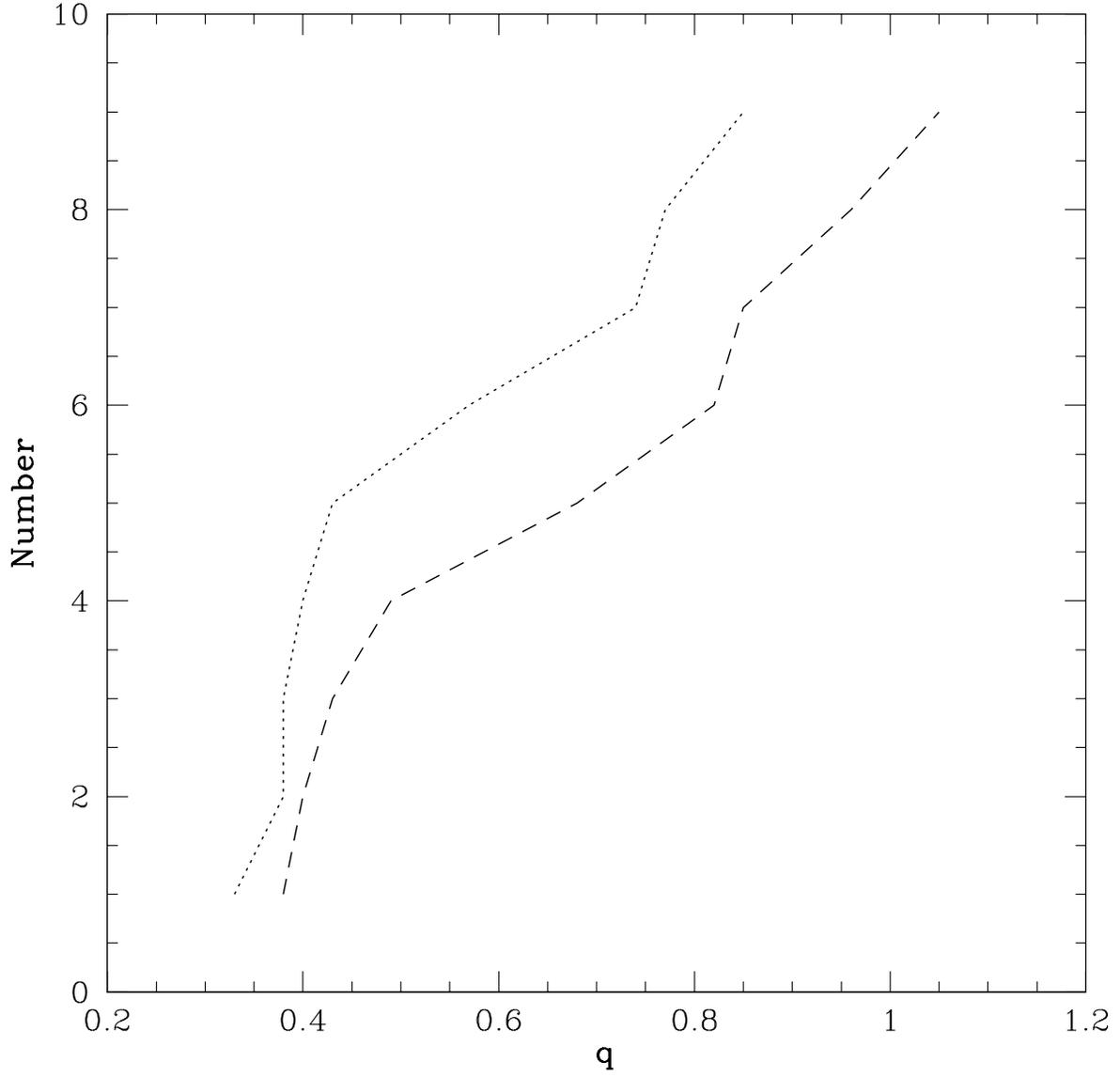}
\caption{The cumulative distribution of mass ratios divided into period groups
 ($\log P <4$:
dashed line; $\log P >4$: dotted  line).  Periods are in days.
 \label{fig13}}
\end{figure}

\begin{figure}
 \includegraphics[width=\columnwidth]{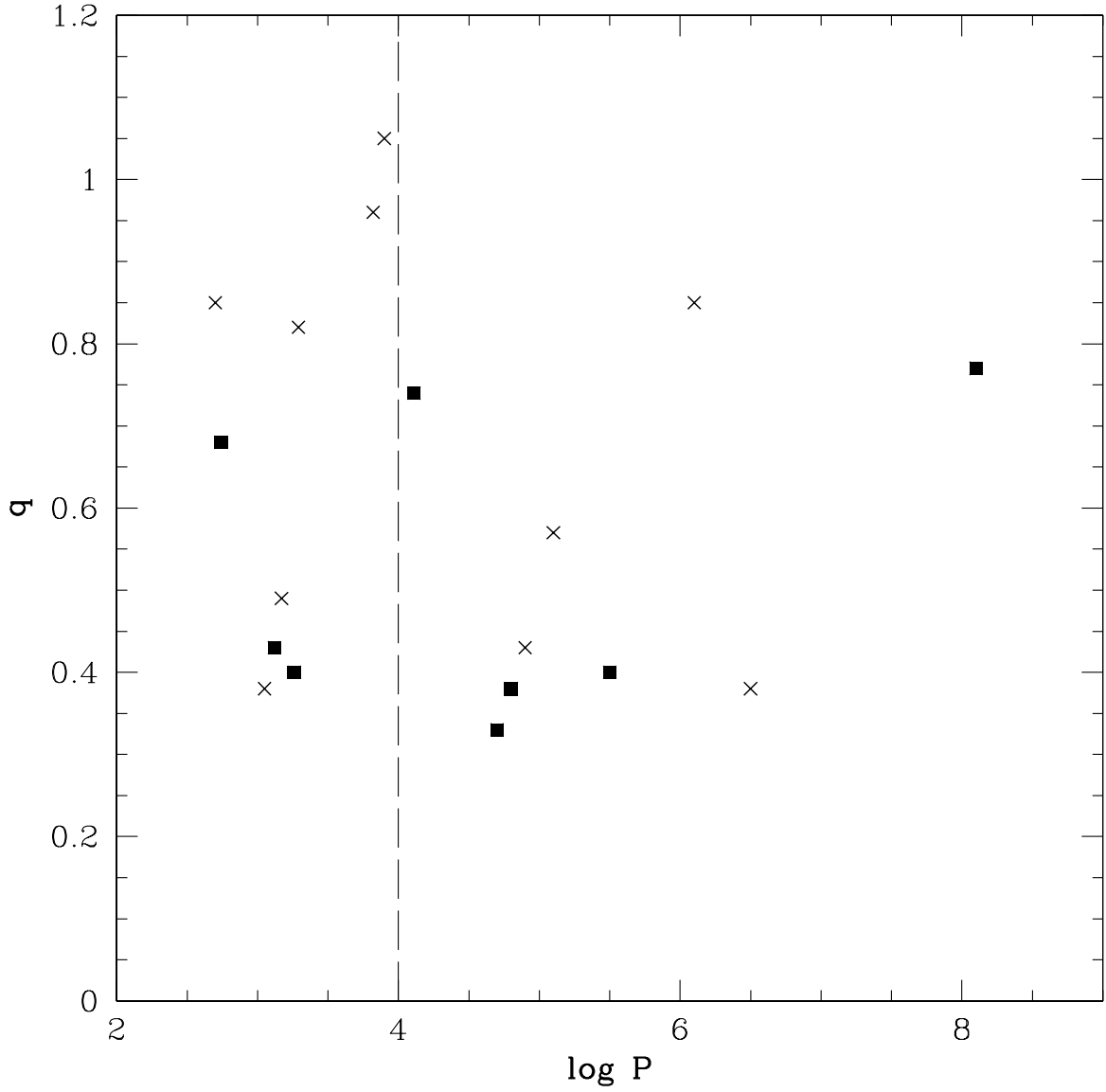}
\caption{The mass ratio $q$ as a function of $\log P$ (in days).  Systems known to be 
triple systems are filled squares; binary systems are x's. The dashed
vertical line
is the  dividing 
line in previous figures at $\log P = 4$ (period in days).
 \label{fig14}}
\end{figure}

\clearpage

\begin{deluxetable}{llccclllccl}
\tabletypesize{\footnotesize}
\tablecaption{Cepheids Brighter than $\langle V\rangle=8$ with 
  Companions M $>$2$M_\odot$\label{}}
\tablewidth{0pt}
\tablehead{
\colhead{Star} & \colhead{Spect.}  
  & \colhead{Ref.} & \colhead{$M_2$}   &
\colhead{Binary}  & \colhead{$P_{\rm orb}$} & 
  \colhead{$\log P_{\rm orb}$} & \colhead{$\log a$} 
  & \colhead{$q$}  & \colhead{Triple?}  & \colhead{Ref.} \\
\colhead{} & \colhead{Type} & \colhead{}  
  & \colhead{[$M_\odot$]}  &
\colhead{Type\tablenotemark{a}}  & \colhead{[days]} & \colhead{[days]} & \colhead{[AU]} & 
  \colhead{[$M_2/M_1$]} & \colhead{}  & \colhead{} \\
}
\startdata
$\eta$ Aql   & B9.8 V & 3  & 2.3 & r & \nodata &  5.5 &      2.3 &   0.40 & t  & 12  \\   
U Aql        & B9.8 V & 1  & 2.3 & o & 1856 &  3.27 &  0.77 &   0.40 & t & 11  \\  
RX Cam       & A0 V   & 1  & 2.2 & o & 1113 & 3.05 & 0.62 &   0.38 &  &   \\         
SU Cas       & B9.5 V & 3  & 2.4 & \nodata  & \nodata&    5.1 &    2.0 &     0.57 &  &   \\        
V659 Cen     & B6.0   & 4  & 4.4 & r & \nodata&  6.1 &    2.7 &    0.85 &  &   \\          
$\delta$ Cep & B7-8   & 10 & 4   & r & \nodata&   8.1 &  4.0 &   0.77  & t & 10 \\
AX Cir       & B6.0   & 4  & 5.0 & o & 6532 &   3.82 & 1.17 &  0.96  &  &  \\
BP Cir       & B6.0   & 4  & 4.7 & om & \nodata  & 3.9 &    1.2 &    1.05  &  &  \\
SU Cyg       & B8.0 V & 1  & 3.2 & o & 549  &   2.74 & 0.42 &  0.68  & t & 11 \\
V1334 Cyg    & B7.0 V & 1  & 4.0 & o & 1938  &   3.29 & 0.80 & 0.82  &  &   \\
T Mon        & A0p    & 8  & 3.0 & om & \nodata    &  4.7 & 1.8 &     0.33  & t & 8 \\
S Mus        & B3 V   & 7  & 5.3 & o & 505 &   2.70   & 0.45 &  0.85  &  &   \\
S Nor        & B9.5 V & 2  & 2.4 & r & \nodata    &    6.5 &  2.9 &     0.38 &  &   \\ 
AW Per       & B6:    & 6  & 4.0 & o & 13100  & 4.12  & 1.36 & 0.74 & t & 11  \\
W Sgr        & A0   V & 1  & 2.2 & r & \nodata  &   4.8 & 1.8 &  0.38  & t &  11 \\
V350 Sgr     & B9.0 V & 5  & 2.5 & o & 1473  &  3.17  & 0.70 & 0.49 &  &   \\
V636 Sco     & B9.5 V & 1  & 2.4 & o & 1318  & 3.12 & 0.67 &  0.43  & t & 11  \\
T Vul        & A0.8 V & 9  & 2.1 & \nodata & \nodata      & 4.9 & 1.8 &        0.43  &  &  \\
\enddata
\tablenotetext{a}{Binary types: o = spectroscopic orbit with known period, given
in col.~6;  om = spectroscopic orbital motion detected, estimated log period
given in col.~7; r = resolved binary, estimated log period given in col.~7}
\tablerefs{
(1) Evans 1995;
(2) Evans 1992c; 
(3) Evans 1991;  
(4) Evans 1994;   
(5) Evans \& Sugars 1997; 
(6) Massa \& Evans 2008;
(7) Evans et al.\ 2006;
(8) Evans et al.\ 1999; 
(9) Evans 1992b;
(10) Benedict et al.\ 2002;
(11) Evans et al.\ 2005;
(12) this paper.
}
\end{deluxetable}

\begin{deluxetable}{lllcc}
\tablecaption{Cepheids with Resolved Companions\label{}}
\tablewidth{0pt}
\tablehead{
\colhead{Star} & \colhead{WFC3}  
  & \colhead{Sep.} & \colhead{Sep.} &  Ref. \\
\colhead{} & \colhead{Obs.\ Date}  
  & \colhead{[$''$]} & \colhead{[AU]} &    \\
}
\startdata
$\eta$ Aql  &  2010 November 20 & 0.66   & 180 &  1  \\
V659 Cen    &  2011 June 5   & 0.63  & 474  &  1 \\
S Nor       &  2011 April 1 & 0.90   & 817  &  1 \\
$\delta$ Cep & \nodata & 40.0 & 10360 &  2 \\
W Sgr & \nodata & 0.16 & 65 &  3 \\
\enddata
\tablerefs{
(1) This paper; (2) Benedict et al.\ 2002; (3) Evans et al.\ 2009
}
\end{deluxetable}

\begin{deluxetable}{lccccccccc}
\tabletypesize{\footnotesize}
\tablecaption{Physical Properties of the Cepheids}
\tablewidth{0pt}
\tablehead{
\colhead{Star}  & \colhead{$P_{\rm puls}$} & \colhead{$M_V$} & \colhead{$E(B-V)$} &    
\colhead{$(B-V)_0$} & \colhead{$\langle V_0\rangle$} & \colhead{BC}  & \colhead{$\log L/L_\odot$} & 
  \colhead{$d$} & \colhead{$M/M_\odot$}     \\
\colhead{}  & \colhead{[days]} & \colhead{[mag]} & \colhead{[mag]}  &   
\colhead{[mag]} & \colhead{[mag]} &  \colhead{[mag]}  & \colhead{} & \colhead{[pc]} & \colhead{}     \\
}
\startdata
$\eta$ Aql    &   7.17  &  $-3.70$  &  0.12  &   0.71 &   3.48 &    $-0.110$ &   3.42 &   273  &   5.7 \\
U Aql         &    7.02  &  $-3.68$ &  0.35  &   0.70  &  5.26  &   $-0.105$  &  3.41  &  614   &   5.7  \\
RX Cam        &  7.91  &  $-3.80$  &  0.63  &   0.61  &  5.47  &   $-0.067$  &  3.44  &  714  &   5.8 \\
SU Cas        &   2.74\rlap{$^a$}  &  $-2.68$  &  0.23  &   0.49 &   5.19  &   $-0.027$ &   2.97  &  375  &  4.2 \\
V659 Cen      & 5.62  &   $-3.44$  &  0.21  &   0.61 &   5.94  &   $-0.067$ &   3.29  &  752  &  5.2 \\
$\delta$ Cep  &  5.36  &   $-3.39$  &  0.09  &   0.57  &  3.68  &   $-0.053$  &  3.27  &  259  &  5.2 \\
AX Cir        &  5.27  &   $-3.37$  &  0.25   &  0.67  &  5.23  &   $-0.091$  &  3.28 &   525  &   5.2 \\
BP Cir        &   3.39\rlap{$^a$} &  $-2.91$  &  0.32  &   0.51  &  6.60  &   $-0.033$  &  3.07 &   798  &   4.5 \\
SU Cyg        &   3.84  & $-3.04$  &  0.08  &   0.56  &  6.62 &    $-0.049$  &  3.13 &   855  &   4.7 \\
V1334 Cyg     & 4.74\rlap{$^a$}  & $-3.26$  &  0.07  &   0.51  &  5.74  &   $-0.033$  &  3.21  &   631  &   4.9 \\
T Mon         &   27.02  &$-5.10$  &  0.14  &   1.09  &  5.66  &   $-0.416$  &  4.10 &  1419  &   9.1 \\
S Mus         &  9.65  &   $-4.01$ &  0.21  &   0.69 &  5.47  &  $-0.100$ &  3.54  &  787  &   6.2 \\
S Nor         &   9.75   & $-4.02$ &  0.19   &  0.77  &  5.77  &   $-0.153$ &   3.56  &  908  &   6.3 \\
AW Per        &   6.46  &  $-3.59$ &  0.53  &   0.57  &  5.72  &   $-0.053$  &  3.35  &  728  &   5.4 \\
W Sgr         &  7.59   &  $-3.76$ &  0.11  &   0.65  &  4.30  &   $-0.083$  &  3.43  &  409   &  5.8 \\
V350 Sgr      &  5.15  & $-3.35$ &  0.32  &   0.55  &  6.41  &   $-0.046$  &  3.25 &   895  &   5.1 \\
V636 Sco      &  6.79  &   $-3.64$ &  0.20  &   0.73  &  5.96  &   $-0.119$  &  3.40  &  832  &   5.6 \\
T Vul         &   4.43  &   $-3.19$  &  0.06  &   0.64  &  5.41  &   $-0.079$ &   3.21  &  561 &   4.9 \\
\enddata
\tablenotetext{a}{First-overtone Cepheid, for which the listed period has been
``fundamentalized'' as discussed in the text}
\end{deluxetable}

\end{document}